\documentclass[useAMS,usenatbib]{mn2e}

\title{The turbulent density  spectrum in the solar wind plasma}
\author[Dastgeer Shaikh]
  {Dastgeer ~Shaikh \thanks{email: dastgeer.shaikh@uah.edu} and G. P. Zank \\
    Department of Physics and Center for Space Plasma and Aeronomic Research (CSPAR)\\
The University of Alabama in Huntsville,
Huntsville. Alabama, 35899}
\date{Received  on April 18, 2009; Revised on Oct 5, 2009}

\pagerange{\pageref{firstpage}--\pageref{lastpage}} \pubyear{2009}

\def\LaTeX{L\kern-.36em\raise.3ex\hbox{a}\kern-.15em
    T\kern-.1667em\lower.7ex\hbox{E}\kern-.125emX}

\newcommand{\be}{\begin{equation}}
\newcommand{\ee}{\end{equation}}

\newcommand{\Fig}[1]{Fig. (\ref{#1})} 
 \newcommand{\eqa}{\begin{eqnarray}}
\newcommand{\eeq}{\end{eqnarray}}

\begin{document}

\label{firstpage}

\maketitle

\begin{abstract}

The density fluctuation spectrum in the solar wind reveals a
Kolmogorov-like scaling with a spectral slope of $-5/3$ in wavenumber
space.  The energy transfer process in the magnetized solar wind,
characterized typically by MHD turbulence, over extended length-scales
remains an unresolved paradox of modern turbulence theories, raising
the question of how a compressible magnetofluid exhibits a turbulent
spectrum that is characteristic of an incompressible hydrodynamic
fluid.  To address these questions, we have undertaken
three-dimensional time dependent numerical simulations of a
compressible magnetohydrodynamic fluid describing super-Alfv\'enic,
supersonic and strongly magnetized plasma fluid. It is shown that a
Kolmogorov-like density spectrum can develop by plasma motions that
are dominated by Alfv\'enic cascades whereas compressive modes are
dissipated.
\end{abstract}


\section{Introduction}
It is a curious observation that electron density fluctuations in the
interstellar medium (ISM) exhibit an omnidirectional Kolmogorov-like
\citep{kol} power spectrum $k^{-5/3}$ (or -11/3 spectral index in
three dimensions) over a 4 to 5 decade range
\citep{amstrong,higdon84,amstrong2}.  The solar wind plasma also
possesses density fluctuations that exhibit a Kolmogorov-like
$k^{-5/3}$ spectrum (Goldstein et al 1995; Matthaeus \& Brown 1988;
Zank \& Matthaeus 1990; Montgomery et al 1987; Lithwick \& Goldreich
2001; Padoan \& Nordlund 1999; Shaikh \& Zank 2007).  Turbulent
processes are believed to be responsible for the observed density
spectrum in the ISM
\cite{amstrong,higdon84,higdon86,amstrong2,elmegreen,scalo1,scalo2}
and solar wind \cite{Goldstein1995, Matthaeus1988, zank90, Montgomery,
  Podesta,Podesta2006,Podesta2008}.  The Kolmogorov-like 5/3 spectrum
is observed in many fluid, space and astrophysical plasmas as
well. For instance, turbulent spectra in the ISM and galaxies are
found to exhibit a Kolmogorov-like scaling in wavenumber space
\cite{roy,Haverkorn,ryu,rickett,Willett,elmegreen2,Dickey,minter}.
Several MHD and hydrodynamic fluid turbulence simulations show a
Kolmogorov-like 5/3 spectrum. Some example of which are
\cite{biskamp,dastgeer2,goldreich,ghosh,kim,kritsuk}. An
exhaustive list of references describing a Kolmogorov-like 5/3
spectrum in hydrodynamic fluid and magnetoplasma turbulence is however
not possible to cite here. Some good reviews such as
\cite{macomb,Lesieur,biskamp,scalo1,scalo2} discuss various possible
scenarios of turbulence in general.  While all of these works show the
possiblity of a self-consistent energy exchange between widely
disparate length-scales in the presence of waves, nonlinear
structures, anisotropy, driving forces etc., the physical processes
leading to a Kolmogorov-like turbulent density fluctuation spectrum is
not yet fully understood.  A great deal of attention has thus focused
on understanding the evolution of MHD turbulence spectra in the
context of the solar wind and ISM
\cite{higdon84,higdon86,Matthaeus1988,zank90,zank93,bayly,Montgomery,goldreich,padoan,dastgeer,dastgeer2}.

Higdon (1984, 1986) interpreted the observed density fluctuations
\citep{amstrong} to be two-dimensional isobaric entropy variations in
which temperature and density gradients are directed oppositely and
both are orthogonal to the local approximately uniform magnetic
field. Based on a pseudosound approximation \citep{light}, Montgomery
et al (1987) related density fluctuations and incompressible
magnetohydrodynamics (MHD) velocity and magnetic field fluctuations.
This approach, called a pseudosound approximation, assumes that
density fluctuations are proportional to the pressure fluctuations
through the square of sound speed. The density perturbations in their
model are therefore âslavedâ to the incompressible magnetic field and
the velocity fluctuations.  This hypothesis was further contrasted by
Bayly et al. (1992) on the basis of their 2D compressible hydrodynamic
simulations by demonstrating that a spectrum for density fluctuations
can arise purely as a result of abondoning a barotropic equation of
state without even requiring a magnetic field. Bayly's (1992) work
ignores magnetic field effects from the outset and hence does not
explain the influence and possible correlation of magnetic field and
corresponding magnetized waves on the density fluctuation
spectrum. The latter has been investigated in the slow solar wind
plasma by Spangler \& Spitler (2004) who suggest that there exists a
strong correlation between the density and magnetic field
fluctuations.  The pseudosound fluid description of compressibility,
justifying the Montgomery et al. (1987) approach to the
density-pressure relationship, was further extended by Matthaeus and
Brown (1988) in the context of a compressible magnetofluid (MHD)
plasma with a polytropic equation of state in the limit of a low
plasma acoustic Mach number (Matthaeus and Brown, 1988). The theory,
originally describing the generation of acoustic density fluctuations
by incompressible hydrodynamics (Lighthill, 1952), is based on a
generalization of Klainerman and Majda's work (Klainerman and Majda,
1981, 1982; Majda, 1984) and accounts for fluctuations associated with
a low turbulent Mach number fluid, unlike purely incompressible
MHD. Such a nontrivial finite departure from the incompressibility
state is termed a 'nearly incompressible' fluid description and is put
forward to provide a possible explanation of the turbulent density
variations that are observed to exhibit a Kolmogorov-like power
spectrum in the solar wind plasma (Montgomery et al. 1987; Matthaeus
\& Brown 1988, Zank \& Matthaeus 1990, 1993, Shaikh \& Zank
2004a,b). In the context of the ISM, a comparative study of
two-dimensional turbulence of self-gravitating supersonic MHD,
hydrodynamic and Burgers turbulence by Scalo et al (1998) suggests a
power law form of density fluctuations that is close to -1.7.  Kim \&
Ryu (2005) report that the slope of the density power spectra in
hydrodynamic turbulence becomes gradually shallower as the rms Mach
number increases and it tends towards a Kolmogorov-like slope when the
rms (or turbulent) Mach number is unity. The high resoultion
simulations of Euler turbulence by Kritsuk et al (2007) suggest that
the inertial range velocity scaling in the strongly compressible
regime (with a spectral index close to -1.95) deviates substantially
from the incompressible Kolmogorov 5/3-law.  Kida \& Orszag (1990)
report that it is only the rotational component of the velocity field
in driven hydrodynamic compressible fluid turbulence that exhibits
spectra very close to that of the incompressible case even for a large
Mach number (close to unity).  Lithwick \& Goldreich (2001) show that
density fluctuations in the ISM plasma are generated by entropy modes
while Alfv\'{e}n waves lead to a Kolmogorov-like spectrum.  This,
however, is not generic to all Alfv\'{e}nic Mach numbers ($M_A$)
\citep{passot}. While both slow and fast modes introduce density
fluctuations at large $M_A$, only the slow mode dominates the
density-magnetic field anticorrelation at relatively small $M_A$
\citep{passot}. However, the slow mode is known to be strongly Landau
damped in a collisionless plasma.  The presence of a mean magnetic
field introduces additional complexities in the energy cascade
processes in MHD turbulence. For instance, the assumption of isotropy
breaks down in the presence the mean or large scale magnetic
field. Along the direction of this large scale magnetic field Alfv\'en
waves suppress the parallel cascade (Iroshnikov 1963; Kraichnan 1965).
The resulting inertial range MHD spectrum is therefore thought to be
flattened from $k^{-5/3}$ to $k^{-3/2}$ (Iroshnikov 1963; Kraichnan
1965).  Within the paradigm of incompressible MHD turbulence,
Goldreich \& Sridhar (1995) proposed that parallel ($k_\parallel$) and
perpendicular ($k_\perp$) modes are correlated by $k_\parallel \propto
k_\perp^{2/3}$ and energy in the perpendicular modes follows a
Kolmogorov-like spectrum when linear (along $B_0$) and nonlinear
(across $B_0$) frequencies balance. By contrast, Shaikh \& Zank (2007)
argue that such balance is not obeyed identically by the entire
inertial range modes, but only by a few modes that critically balance
linear and nonlinear Alfv\'enic frequency. Boldyrev (2005) proposes a
scale dependent anisotropic power law that appears to differ from the
in-situ solar wind observations (Podesta et al 2008).  While in-situ
spacecraft observations of magnetic field fluctuations in solar wind
plasma are shown to follow the Kolmogorov-like $k^{-5/3}$ spectrum
(Montgomery et al 1987; Zhou et al 1990; Goldstein et al 1995), the
velocity field tends to show a close consistency with a $k^{-3/2}$
spectrum (Podesta et al 2006, 2007). The latter is contrasted by
Roberts (2007) that the magnetic field and velocity in the solar wind
do not evolve in the same way with helocentric distance. Based on
Voyager observations, Roberts (2007a,b) argues that velocity spectrum
relaxes towards a likely asymptotic state through spectral steepening
and acquires a spectral index of -5/3, finally mathching the magnetic
field spectrum. Roberts (2007b) further argues that -3/2 is accidental
and transient, and that the -5/3 slope is the eventual state of all
the fluctuations. In a comprehensive review, Bruno and Carbone (2005)
and Veltri (1980) describe that low frequency solar wind velocity
fluctuations closely follow a Kolmogorov-like spectrum, and the
intermediate region (between high and low frequency) do not allow us
to distinguish between a Kolmogorov spectrum (-5/3) and a Kraichnan
spectrum (-3/2). Bavassano et al (2005) describe that large scale
velocity field fluctuations in the polar solar wind closely follow a
Kolmogorov-like 5/3 spectrum.

The discrepancy in the magnetic and velocity field spectra continues
to be an unresolved issue.  Since our simulations tend to favor a
Kolmogorov-like $k^{-5/3}$ velocity spectrum in which density spectrum
($k^{-5/3}$) closely follows the velocity spectrum through a passive
convection, we support a $k^{-5/3}$ Kolmogorov-like spectrum for the
velocity field fluctuations in solar wind. What is notable in our work
is the dissipation of the high frequency component due to the damping
(described in Section 2) of compressible plasma motion that suppresses
the small scale and high frequency compressive turbulent modes. The
MHD plasma therefore evolves towards a nearly incompressible state and
the density is convected passively by the velocity field to yield a
$k^{-5/3}$ spectrum.

Perhaps, the most striking point about the Kolmogorov-like $k^{-5/3}$
spectrum is its ubiquitous persistence in fluids and plasmas
regardless of whether they are (in)compressible, (un)magnetized,
(an)isotropic and (un)driven. The observed Kolmogorov-like density
spectrum yields two paradoxes; (1) why does a compressible magnetized
fluid behaves as though it were incompressible and umagnetized, and
(2) Why do the density fluctuations, an apparently quintessential
compressive characteristic of magnetized turbulence, yield a
Kolmogorov power law spectrum characteristic of incompressible
hydrodynamic turbulence?  These questions have to be answered if we
are to address the origin of the 5/3 spectrum in MHD turbulence in
general and the ISM density power law spectrum, in particular.  In
this paper, we address these issues within the context of fully 3D
simulations of compressible, anisotropic, driven Magnetohydrodynamics
(MHD) turbulence to understand how and why a supersonic, super
Alfv\'enic, and low plasma $\beta$ ($\beta$ the ratio of plasma
pressure and magnetic pressure) compressible MHD fluid should exhibit
a Kolmogorov-like wavenumber spectrum in density. We find a strong
correlation between the intrinsic MHD waves (i.e. Alfv\'en, fast \&
slow magnetoacoustic modes) and nonlinear inertial range turbulent
cascades that suggests that nonlinear mode coupling interactions in
compressible MHD turbulence tend to dissipate high frequency
magnetoacoustic modes. Consquently, the inertial range cascade is
governed predominantly by Alfv\'enic interactions that passively
convect the density field to yield a Kolmogorov-like $k^{-5/3}$
spectrum.

In section 2, we describe the governing equations, and critical
assumptions of our 3D magnetohydrodynamic simulations. Section 3
describes results from nonlinear fluid simulations of compressible,
driven, anisotropic, homogeneous, turbulent magnetohydrodynamic (MHD)
plasma. Our simulations demonstrate that density, magnetic and
velocity fields in MHD turbulence follow an omnidirectional
Kolmogorov-like $k^{-5/3}$ turbulent spectrum. The physical arguments
explaining the evolution of a $k^{-5/3}$ spectrum are described in
section 4 that deal primarily with turbulent damping of non-solenoidal
velocity field fluctuations. In section 5, we outline our results for
anisotropic cascades that result from the presence of a mean magnetic
field in MHD turbulence. Finally, section 6 summarizes our major
results.

\section{MHD model}

Our underlying magnetohydrodynamic (MHD) model assumes that
characteristic fluctuations in the magnetofluid plasma are initially
isotropic, homogeneous, thermally equilibrated and turbulent. A large
scale constant magnetic field is present and drives anisotropic
turbulent cascades in an initially isotropic spectral distribution in
compressible MHD turbulence.  The characteristic turbulent
fluctuations in the plasma are assumed in our model to be much bigger
than shocks or discontinuities. In other words, we ignore the
influence of shock on turbulent fluctuations. Our work incorporating
the effect of shocks on turbulent spectra is initiated in
\cite{zank2007,zank2006}.  The boundary conditions are periodic, hence
mode coupling interactions in the local spatial region are considered.

The fluid model describing nonlinear turbulent processes in the
magnetofluid plasma, in the presence of a background magnetic field,
can be cast into plasma density ($\rho_p$), velocity (${\bf U}_p$),
magnetic field (${\bf B}$), pressure ($P_p$) components according to
the conservative form
\be
\label{mhd}
 \frac{\partial {\bf F}_p}{\partial t} + \nabla \cdot {\bf Q}_p={\cal Q},
\ee
where,
\[{\bf F}_p=
\left[ 
\begin{array}{c}
\rho_p  \\
\rho_p {\bf U}_p  \\
{\bf B} \\
e_p
  \end{array}
\right], 
{\bf Q}_p=
\left[ 
\begin{array}{c}
\rho_p {\bf U}_p  \\
\rho_p {\bf U}_p {\bf U}_p+ \frac{P_p}{\gamma-1}+\frac{B^2}{8\pi}-{\bf B}{\bf B} \\
{\bf U}_p{\bf B} -{\bf B}{\bf U}_p\\
e_p{\bf U}_p
-{\bf B}({\bf U}_p \cdot {\bf B})
  \end{array}
\right],\]
\[{\cal Q}=
\left[ 
\begin{array}{c}
0  \\
{\bf f}_M({\bf r},t) +\mu \nabla^2 {\bf U}+\eta \nabla (\nabla\cdot {\bf U})  \\
\eta \nabla^2 {\bf B}  \\
0
  \end{array}
\right]
\] 
and
\[ e_p=\frac{1}{2}\rho_p U_p^2 + \frac{P_p}{\gamma-1}+\frac{B^2}{8\pi}.\]
Equations (1) are normalized by typical length $\ell_0$ and time $t_0
= \ell_0/V_A$ scales in our simulations such that
$\bar{\nabla}=\ell_0{\nabla}, \partial/\partial
\bar{t}=t_0\partial/\partial t, \bar{\bf U}_p={\bf U}_p/V_A,\bar{\bf
  B} ={\bf B}/V_A(4\pi \rho_0)^{1/2}, \bar{P}=P/\rho_0V_A^2,
\bar{e}_p=e_p/\rho_0V_A^2, \bar{\rho}=\rho/\rho_0$. The bars are
removed from the normalized equations (1). $V_A=B_0/(4\pi \rho_0)^{1/2}$
is the Alfv\'en speed

The rhs in the momentum equation denotes a forcing functions (${\bf
  f}_M({\bf r},t)$) that essentially influences the plasma momentum at
the larger length scale in our simulation model.  With the help of
this function, we drive energy in the large scale eddies to sustain
the magnetized turbulent interactions. In the absence of forcing, the
turbulence continues to decay freely.  While the driving term modifies
the momentum of plasma, we conserve density (since we neglect
photoionization and recombination). The large-scale random driving of
turbulence can correspond to external forces or instabilities for
example fast and slow streams, merged interaction region etc in the
solar wind, supernova explosions, stellar winds in the ISM, etc. The
magnetic field evolution is governed by the usual induction equation,
i.e. Eq. (\ref{mhd}), and obeys the frozen-in-field theorem unless
dissipative mechanism introduces small-scale damping.  Note carefully
that MHD plasma momentrum equation contains dissipative terms on the
rhs.  It is the term in the momentum equation (i.e. $\partial (\rho_p
{\bf U}_p)/\partial t \cdots $) that is proportional to $\mu \nabla^2
{\bf U}_p +\eta \nabla (\nabla \cdot {\bf U}_p)$. The latter (along
with the other terms in the equation) is divided by the density
$\rho_p$ field during the evolution to calculate the velocity
field.

\section{Nonlinear Simulations}

We have developed a fully three dimensional compressible MHD code to
study the nonlinear mode coupling interaction in the context of
compressible MHD turbulence.  Details of our code are described in
\citep{dastgeer,dastgeer2}. In the simulations, all the fluctuations
are initialized isotropically with random phases and amplitudes in
Fourier space and an initial spectral shape close to $k^{-2}$ ($k$ is
the Fourier mode, which is normalized to the characteristic turbulent
length-scale $l_0$). We have carried out the simulations for both
decaying and driven-dissipative cases with and without external
magnetic field $B_0$. Since we are interested in a local region of the
solar wind magnetofluid plasma, the computational domain employs a
normalized three-dimensional periodic box of volume $\pi^3$. Other
parameters are $\gamma=5/3, \beta=0.1-2.0, M_A=1.0-2.0, M_s=1.0-2.0,
\eta=\mu=10^{-14}-10^{-15}, k_f<15.0$, where $\gamma, M_A, M_s, \eta,
\mu$ and $k_f$ are respectively the ratio of specific heats, Alfv\'en
Mach number, sonic Mach number, viscosity, magnetic diffusion and
energy injection modes.  Our MHD model does not include
(photo)ionization and radiation terms. Hence source or sink terms
corresponding to the density fluctuations are not included in our MHD
model. Additionally, the localized dissipation is effective in our
simulations at the small-scales where it damps the plasma motions. The
small scale dissipation in the local interstellar medium or solar wind
may result from e.g. radiative cooling or ion-neutral collisions
\citep{spangler}. Accordingly, the small-scale dissipation in our
model corresponds to collisional or viscous effects and is associated
with the small-scale damping that is responsible only for the cascade
of large-scale energy into the smaller scales thereby producing a
well-defined inertial range turbulent spectrum.  By contrast, the
large-scales and the inertial range turbulent fluctuations remain
unaffected by direct dissipation of the smaller scales.

The initial kinetic and magnetic energies are equi-partitioned between
the velocity and the magnetic fields. The latter helps treat the
transverse or shear Alfv\'en and fast/slow magnetosonic waves on an
equal footing, at least during the early phase of the simulations.
Magnetoplasma turbulence evolves under the action of nonlinear
interactions in that larger eddies transfer their energy to smaller
ones through a forward cascade. According to \citep{kol,iros,krai},
the cascade of spectral energy is mediated by a local interaction
amongst the neighboring Fourier modes that continues until the energy
in the smallest turbulent eddies is dissipated due to the finite
Reynolds number. This leads to the damping of small scale motions as
well. This results in a net decay of the turbulent sonic Mach number
$M_s$ associated with the large scale fluctuations. If turbulence is
not driven at large scales, the turbulent sonic Mach number continues
to decay from a supersonic ($\tilde{M}_s>1$) to a subsonic
($\tilde{M}_s<1$) regime \citep{dastgeer}. This indicates that
nonlinear cascades predominantly cause supersonic MHD plasma
fluctuations to become subsonic.  In our decaying turbulence
simulation, the large-scale energy simply migrates towards the smaller
scales by virtue of nonlinear cascades in the inertial range and is
dissipated at the smallest turbulent length-scales.  On the other
hand, spectral transfer in driven turbulence follows a similar cascade
process as in the decaying turbulence case. However, the inertial
range spectrum in the latter is maintained by a large scale forcing at
$k<5$.  The spectral transfer of turbulent energy in the neighboring
Fourier modes in globally isotropic and homogeneous hydrodynamic and
magnetohydrodynamic turbulence is the widely accepted paradigm
\citep{kol} that leads to Kolmogorov-like energy spectra, while
\citep{iros, krai} describe turbulent spectra in the presence of a
mean or local $B_0$.  The most striking effect, however, to emerge
from the decay of the turbulent sonic Mach number is that the density
fluctuations begin to scale quadratically with the subsonic turbulent
Mach number as soon as the compressive plasma enters the subsonic
regime, i.e. $ \delta \rho \sim {\cal O}(\tilde{M}_s^2)$ when
$\tilde{M}_s<1$. This was demonstrated in our 3D simulations of a
compressible MHD plasma \citep{dastgeer}.  It signifies an essentially
{\it weak} compressibility in the magnetoplasma, and is consistent
with a {\it nearly incompressible} state
\citep{Matthaeus1988,zank90,zank93, dastgeer}.

\begin{figure}
\vspace{200pt}
\begin{center}
\end{center}
\caption{\label{fig1} (Left) Velocity fluctuations are dominated by
  shear Alfv\'enic motion and thus exhibit a Kolmogorov-like
  $k^{-5/3}$ spectrum. The middle curve shows the magnetic field
  spectrum. Density fluctuations are passively convected by the nearly
  incompressible shear Alfv\'enic motion and follow a similar spectrum
  in the inertial range. The numerical resolution in 3D is
  $512^3$. (Right) The evolution of Alfv\'enic ($k_A$) and fast/slow
  magnetosonic ($k_{MS}$) modes demonstrates that the spectral
  cascades are dominated by Alfv\'enic modes.}
\end{figure}

In the context of the magnetoplasma being nearly incompressible, the
density fluctations exhibit a weak compressibility in the gas and are
convected predominantly passively in the background incompressible
fluid flow field. This hypothesis can be verified straightforwardly by
investigating the density spectrum which should be slaved to the
incompressible velocity spectrum. This is shown in \Fig{fig1} which
illustrates that the density fluctuations follow the velocity
fluctuations in the inertial regime over the long time (several
Alfv\'en transit time) evolution of MHD turbulence.  The evolution of
compressible magnetoplasma from a(n) (initial) supersonic to a
subsonic or nearly incompressible regime is gradual and it results in
the density field following the velocity fluctuations. In the subsonic
regime, compressibility weakens substantially so that density
fluctuations are advected only passively.  A passively convected fluid
exhibits a similar inertial range spectra as that of its background
flow field \citep{macomb}. Likewise, subsonic density fluctuations in
our simulations exhibit a Kolmogorov-like $k^{-5/3}$ spectrum similar
to the background velocity fluctuations in the inertial range.  This,
we believe, provides a plausible explanation for the Kolmogorov-like
density spectrum observed in MHD turbulence i.e. they are convected
passively in a field of nearly incompressible velocity fluctuations
and acquire identical spectral features [as shown in \Fig{fig1}].  The
passive scalar evolution of the density fluctuations is associated
essentially with incompressiblity and can be understood directly from
the continuity equation as follows. Expressing the fluid continuity
equation as $(\partial_t + {\bf U} \cdot \nabla) \ln \rho = - \nabla
\cdot {\bf U} $, where the rhs represents compressiblity of the
velocity fluctuations, shows that the density field is advected
passively when the velocity field of the fluid is nearly
incompressible with $\nabla \cdot {\bf U} \simeq 0$.  The theoretical
basis illustrating the nonlinear damping of the non-solenoidal
component of velocity field, i.e. $\nabla \cdot {\bf U} \rightarrow
0$, is described {\em quantitatively} in the following section.

\section{Turbulent evolution of non-solenoidal velocity field}

Understanding, how an initially non-solenoidal velocity field evolves
towards a solenoidal field is important as it explains the evolution
of a compressible MHD magnetoplasma from a supersonic to a subsonic or
nearly incompressible state that yields a passively advected
Kolmogorov-like $k^{-5/3}$ density spectrum.  Any given velocity field
can be decomposed into solenoidal and non-solenoidal components or,
equivalently, longitudinal and transverse components, which we shall
here refer to as fast/slow magnetosonic and Alfven components,
respectively. The simulations show that the amplitudes of fast/slow
magnetosonic components described by the quantity $k_{MS}$ decay more
rapidly than the amplitudes of the Alfvenic components.  Consequently,
they will be inefficient in cascading the corresponding inertial range
spectral energy. Hence nonlinear interactions, in the inertial range,
are governed predominantly by non-dissipative Alfv\'enic modes ($k_A$)
that survive collisional damping in compressible MHD turbulence.  This
is quantitatively demonstrated in \Fig{fig1}[right panel].

The damping of a non-solenoidal velocity component, in part, explains
the observed Kolmogorov-like $k^{-5/3}$ in our simulations.  To this
end, it is essential to distinguish the Alfv\'enic and non-Alfv\'enic,
i.e. corresponding to the compressional or due to slow and fast
magnetosonic modes, contributions to the turbulent velocity
fluctuations.  To identify the distinctive role of Alfv\'enic and
fast/slow (or compressional) MHD modes, we introduce diagnostics that
distinguish the energies corresponding to Alfv\'enic and slow/fast
magnetosonic modes.  Since the Alfv\'enic fluctuations are transverse,
the propagation wave vector is orthogonal to the velocity field
fluctuations i.e.  ${\bf k} \perp {\bf U}$, and the average spectral
energy contained in these (shear Alfv\'enic modes) fluctuations can be
computed as
\[ \langle k_{A} (t) \rangle \simeq \sqrt{\frac{{\sum_{\bf k}}|i {\bf k} \times {\bf U}_{\bf k}|^2}{\sum_{\bf k} |{{\bf U}_{\bf k}}|^2}}.\]
The above relationship leads to a finite spectral contribution from
the $| {\bf k} \times {\bf U}_{\bf k}|$ characteristic turbulent
Alfv\'enic modes.  On the other hand, fast/slow magnetosonic modes
propagate longitudinally along the velocity field fluctuations, i.e.
${\bf k} \parallel {\bf U}$ and thus carry a finite component of
energy corresponding {\em only} to the $i {\bf k} \cdot {\bf U}_{\bf
  k}$ part of the velocity field, which can be determined from the
following relationship
\[ \langle k_{MS} (t) \rangle \simeq \sqrt{\frac{{\sum_{\bf k}}|i {\bf k} \cdot {\bf U}_{\bf k}|^2}{\sum_{\bf k} |{{\bf U}_{\bf k}}|^2}}.\]
The expression of $k_{MS}$ essentially describes the modal energy
contained in the non-solenoidal component of the MHD turbulent modes.

\begin{figure}
\vspace{200pt}
\begin{center}
\end{center}
\caption{\label{fig2} Turbulent energy associated with the
  characteristic inertial range modes corresponding to $| {\bf k}
  \times {\bf U}_{\bf k}|$ and ${\bf k} \parallel {\bf U}$
  fluctuations is shown. Our simulations show that the velocity
  fluctuations are dominated by shear Alfv\'enic motion whose
  contribution, corresponding to the curve represented by $|U(k)_{\rm
    Curl U}|^2$, is more than an order larger than that corresponding
  to the fast/slow magnetosonic modes and is shown by $|U(k)_{\rm Div
    U}|^2$. This result is consistent with Fig 1 (right panel) that
  depicts the mode coupling evolution of the two MHD modes.}
\end{figure}

The quantative evolution of the characteristic modes corresponding to
the Alfv\'enic $k_A$ and slow/fast compressional magnetosonic $k_{MS}$
modes is depicted in Fig. 1 (right panel). Although the modal energies
in $k_A$ and $k_{MS}$ modes are identical initially, a disparity in
the cascade rate develops, and the energy in longitudinal (or
compressional) fluctuations associated with the non-solenoidal
velocity field decays far more rapidly than the energy in the
Alfv\'enic modes. The Alfv\'enic modes, after a modest initial decay,
sustain the energy cascade processes by actively transferring spectral
power amongst various Fourier modes. By contrast, the fast/slow
magnetosonic modes ($k_{MS}$) progressively weaken and suppress their
corresponding spectral contribution in the turbulent energy
cascades. The difference in the cascades corresponding to $k_{A}$ and
$k_{MS}$ modes persists even at long times. The $k_{MS}$ mode
represents collectively a dynamical evolution of small-scale fast plus
slow magnetosonic cascades. The physical implication, however, that
emerges from Fig. 1 is that the fast/slow magnetosonic modes
($k_{MS}$) do not contribute efficiently to the spectral transfer
process, and that the cascades are governed predominantly by
non-dissipative Alfv\'enic modes that survive the collisional damping
in compressible MHD turbulence.  Correspondingly, the turbulent energy
associated with the Alfv\'enic modes makes a dominant contribution to
the velocity fluctuation spectrum when compared to the magnetosonic
modes. We have made measurements of the turbulent energy in the
Alfv\'enic and fast/slow magnetosonic modes to quantify their
respective contributions to the velocity field fluctuation
spectrum. This is shown in Fig. (2). In Fig (2), $|U(k)_{\rm Curl
  U}|^2 = \sum_k | {\bf k} \times {\bf U}_{\bf k}|^2/ \sum_k { k^2}$
and $|U(k)_{\rm Div U}|^2 = \sum_k | {\bf k} \cdot {\bf U}_{\bf k}|^2/
\sum_k { k^2}$.  Clearly, the energy contribution by Alfv\'enic modes
(parallel to $B_0$ and orthogonal to the velocity field) is more than
10 times that of the fast/slow magnetosonic modes (parallel to the
velocity field). This clarifies that it is the predominance of
Alfv\'enic modes (Fig (1) \& (2)) in inertial range cascades that
primarily lead to a Kolmogorov-like spectrum.  Damping of Alfv\'en
waves is possible by ion-neutral collisions as pointed out by
\citep{kulsrud,balsara} in the context of molecular clouds [for more
  references, see \citep{dastgeer3}].  In our present simulations, we
do not include ion-neutral damping as we focus mainly on the solar
wind plasma.  This nonetheless suggests that because of the decay of
the fast/slow magnetosonic modes in compressible MHD plasmas,
supersonic turbulent motions become dominated by subsonic motions and
the nonlinear interactions are sustained primarily by Alfv\'enic modes
thereafter; the latter being incompressible. One of the implications
of the turbulent damping of non-solenoidal velocity field ($i {\bf k}
\cdot {\bf U}_{\bf k}$) in an MHD fluid is that compressible modes
make negligible or no contribution to the inertial range energy
cascade. The cascade is thus determined primarily by the
incompressible Alfv\'en modes that passively convect the density
fluctuations. This point is further consistent with the MHD fluid
continuity equation which in $k$ space reads as follows.
\[\left( \frac{\partial }{\partial t} + i\sum_{\bf k} \delta({\bf k}+{\bf k'}) {\bf U}_{\bf k} \cdot {\bf k'} \right) 
 \ln \rho({\bf k},t) \simeq -i {\bf k} \cdot {\bf U}_{\bf k}.\] The
 nonlinear mode coupling interations associated with the Dirac delta
 function $\delta$ are finite only for those interactions that obey
 the Fourier diad ${\bf k}+{\bf k'}=0$ in spectral space.  It follows
 from our simulations that the turbulent damping of the non-solenoidal
 velocity field on the rhs of the continuity equation makes an
 insignificant contribution to the inertial range energy cascade. The
 nonlinear mode coupling interactions in MHD turbulence are therefore
 dominated by convective transport that leads to a passive convection
 of density fluctuations. The density fluctuations subsequently follow
 the inertial range spectrum and are identical to the background
 nearly incompressible velocity field and thus have a Kolmogorov-like
 $k^{-5/3}$ spectrum \cite{kol,macomb,Lesieur} in MHD turbulence. Our
 simulation results described in Fig. (1) are fully consistent with
 this scenario.

\begin{figure}
\vspace{200pt}
\begin{center}
\end{center}
\caption{\label{fig3} (Left) Evolution of $k_\parallel/k_\perp$ in
  anisotropic magnetofluid turbulence shows that $k_\perp$ becomes
  progressively dominant. (Right) Anisotropic magnetic field spectra
  along and across the $B_0$ are consistent with the left panel. The
  spectra are computed in the steady state (close to $11-13 ~l_0/v_0$)
  and are from the same simulation. The forcing mode spans the
  wavenumber band $3<k_f<5 $.}
\end{figure}

\begin{figure}
\vspace{200pt}
\begin{center}
\end{center}
\caption{\label{fig4} Spectral distribution  of the ratio
$|B(k_\perp)|^2/|B(k_{\parallel})|^2$.}
\end{figure}

\section{Anisotropic turbulent cascades}
We find that the presence of a large scale magnetic field $B_0$ (along
the $\hat{z}$-direction) introduces anisotropy in the distribution of
energy in wavevector space such that the rms wavenumbers along
($k_\parallel$) and across ($k_\perp$) the mean magnetic field $B_0$
show a discrepancy i.e. $k_\parallel \ne k_\perp$ [see \Fig{fig3}]. We
employ the following diagnostics to monitor the evolution of the rms
wavenumbers $k_\perp$ and $k_\parallel$ in time.  The rms $k_\perp$
mode is determined by averaging over the entire turbulent spectrum
weighted by $k_\perp$, thus
\[ k_{\perp} = \langle k_\perp(t)\rangle = \sqrt{\frac{\sum_k |k_\perp B(k,t)|^2}{\sum_k |B(k,t)|^2}}. \]
Here $\langle \cdots \rangle$ represents an average over the entire
Fourier spectrum, $ k_\perp=\sqrt{k_x^2+k_y^2}$.  Similarly, the
evolution of the $k_\parallel = k_z$ (along the $B_0$ direction) mode
is determined by the following relation,
\[ k_{\parallel} = \langle k_\parallel(t) \rangle= \sqrt{\frac{\sum_k |k_\parallel B(k,t)|^2}{\sum_k |B(k,t)|^2}}. \]
The wavenumbers $k_x, k_y$ and $k_z$ are respectively along the $x, y$
and $z$ directions.  It is clear from these expressions that the
$k_\perp$ and $k_\parallel$ modes exhibit isotropy when $k_\perp
\simeq k_\parallel$. Any deviation from this equality corresponds to
spectral anisotropy.  We follow the evolution of $k_\perp$ and
$k_\parallel$ in our simulations. We find a disparity in the magnetic
field fluctuation spectrum along 
\[|B(k_{\parallel})|^2 =
\sum_{k_{\perp} }
|B(k_\perp,k_\parallel)|^2 dk_{\perp}, \]
and across 
\[|B(k_{\perp})|^2
= \sum_{k_{\parallel} } |B(k_\perp,k_\parallel)|^2
dk_{\parallel}\] 
the mean magnetic field (see Fig. 3, right panel).  Note that the
discrete summations $\sum_{k_{\parallel}}$ and $\sum_{k_{\perp} }$ are
carried over $k_\parallel$ and $k_\perp$ modes respectively.  The
presence of the mean magnetic field inhibits turbulent cascades in the
parallel (to $B_0$) direction and hence the characteristic modes
($k_\parallel$) along the mean magnetic field are suppressed, while
the modes in the orthogonal direction remain unaffected.  The
evolution in Fig. (3) therefore shows that the initial isotropic ratio
$k_\parallel/k_\perp \approx 1$ progressively evolves towards
anisotropy $k_\parallel/k_\perp < 1$.

The suppression of the $k_\parallel$ mode is caused by the excitation
of Alfv\'en waves, which act to weaken spectral transfer along the
direction of propagation. This can be understood as follows; We assume
that the spectral transfer, mediated by propagating Alfv\'en waves,
can be described by a three wave interaction mechanism, for which the
frequency and wavenumber resonance criteria are, respectively,
expressed by \cite{sheb,Matthaeus1998}
\[ \pm \omega_3 = \omega_1 - \omega_2, \]
and
\[ {\bf k}_3 = {\bf k}_1 +{\bf k}_2. \]
The frequency-wavenumber resonance conditions indicate that two
Alfv\'en waves $(\omega_1,{\bf k}_1)$ and $(\omega_2,{\bf k}_2)$
mutually interact and give rise to the third wave $(\omega_3,{\bf
  k}_3)$. Such conditions could, in principle, hold for a set of
infinite waves as the indices `$1$' and `$2$' are merely dummy
indices. With the help of the Alfv\'en wave dispersion relation
($\omega={\bf k} \cdot {\bf B}_0$) and the component of wavenumber
matching relation along $B_0$, i.e.  $k_{3_\parallel}=k_{1_\parallel}+
k_{2_\parallel}$, we infer that either the $k_{1_\parallel}=0$ or
$k_{2_\parallel}=0$.  It follows from the frequency-wavenumber
resonance conditions that either the $k_{1_\parallel}=0$ or
$k_{2_\parallel}=0$. Owing to the absence of one of the parallel (to
$B_0$) components of the modes, nonlinear mode coupling interaction
becomes inefficient in transferring the inertial range spectral energy
in the parallel direction. Hence there is very little cascading along
the magnetic field direction.  Thus, the parallel wavenumbers
($k_\parallel$) appear to be suppressed and the spectral cascade
mainly occurs in the perpendicular wavenumbers
($k_\perp$). Consequently, the magnetic field spectrum along the mean
$B_0$ is depleted.  This, we suggest, explains the wavenumber
disparity $k_\parallel \ne k_\perp$ observed in our simulations [see
  Fig 3].  While the turbulent cascades are effected locally by the
presence of the mean magnetic field, the 3D volume averaged inertial
range spectra in our simulations continue to exhibit a power law close
to that of \Fig{fig1}.

The question of anisotropic inertial range cascades has long been
debated in the context of MHD turbulence. It dates back to the seminal
work of Iroshnikov \cite{iros} and Kraichnan (1965), who first pointed
out that the presence of a large-scale or self-consistently generated
magnetic field influences the spectral power cascade mechanism in a
complicated manner \citep{iros,krai}.  Kraichnan (1965) addressed the
interaction of magnetized turbulent eddies with Alfv\'en waves,
excited as a result of a mean magnetic field. Owing to the presence of
waves in a MHD fluid, turbulent correlations between velocity and
magnetic field and the corresponding energy transfer time are
determined primarily by $\tau \sim (b_0 k)^{-1}$, where $b_0$ is a
typical amplitude of a local magnetic field and is dimensionally
identical to that of the velocity field by virtue of Els$\ddot{\rm
  a}$sser's symmetry (Kraichnan 1965). According to Kraichnan (1965),
it is this time scale that leads to the modification in the energy
transfer (or cascade) because of the wave-turbulent eddy interaction
process and without this modification the energy cascade rates would
be determined by typical hydrodynamic eddy interaction time.  The
assumption of isotropy in the Kolmogorov energy spectrum \cite{kol}
was later modified to include the mean or large scale magnetic field
\citep{krai,sheb,higdon84,higdon86,gs95,Matthaeus1998}, in part,
because the presence of waves in MHD turbulence can significantly
influence the energy cascade dynamics in the wavenumber space.  An
attractive analysis was presented by Shebalin et al \citep{sheb}, who
demonstrated that the presence of a mean (or dc) magnetic field
introduces a spectral anisotropy in the MHD turbulent spectrum.  The
observed anisotropic cascade in their work was understood to be due to
the presence of Alfv\'{e}n waves, which are excited and propagate
along the mean magnetic field.  The non-dispersive propagating
Alfv\'{e}n waves in the presence of a mean magnetic field give rise to
distinct energy cascade rates along and across the large-scale
magnetic field, thereby leading to an asymmetry in the spectral
transfer rates. This can be understood as follows; The inertial range
energy cascade rates depend on wavenumber (or modes). Since the
presence of the background magnetic field depletes the parallel mode
(i.e. $k_\parallel$ or $k_z$) and leaves the perpendicular mode
$k_\perp$ unaffected, the spectral energy transfer corresponding to
the $k_\parallel$ mode is suppressed in the direction of the mean or
background magnetic field. By contrast, the perpendicular transfer of
spectral energy remains unaffected. This is quantified by Fig (3),
right panel in which the spectrum of $|B(k_\parallel)|^2$ is
suppressed as compared to that of $|B(k_\perp)|^2$. It is because of
this disparity, the energy cascades along and across the background
magnetic field are distinct.
Such asymmetric cascades, in agreement with arguments
based on the frequency-wavenumber resonance conditions described as
above, produce spectral anisotropy in MHD turbulence
\citep{sheb,Matthaeus1998}.  In the context of the solar wind plasma,
fast ($>500$km/s) and slow ($<400$km/s) streams are dominated
respectively by $k_\parallel$ and $k_\perp$ anisotropic modes
\cite{dasso}.  A more quantitative treatment of anisotropic cascade
was put forward by Goldriech and Sridhar \citep{gs95} for MHD
turbulence, suggesting that the Kolmogorov energy spectrum corresponds
to wavenumbers perpendicular to the local magnetic field so that
$E(k_\perp) \sim k_\perp^{-5/3}$ (where $k_\perp$ is wavenumber
perpendicular to the local magnetic field), whereas the parallel wave
number scales as $k_\parallel \propto k_\perp^{2/3}$.  Boldyrev (2005)
proposes that a Goldreich-Sridhar like spectrum \cite{gs95} results
from a weak large-scale magnetic field, while the limit of strong
anisotropy, that is, strong large-scale magnetic field, corresponds to
the Iroshnikov-Kraichnan \cite{iros,krai} scaling of the spectrum.
This suggests that there exists an asymmetry in the energy cascade
mechanism, which is believed to be due primarily to the presence of
large-scale local magnetic fields in the turbulent MHD flow. Our
simulation results describing the anisotropic MHD turbulent cascades,
depicted in Fig. (3), are further consistent with \citep{sheb}.

\section{Summary}
Our work proposes a self-consistent physical paradigm for the
development of the density fluctuations spectrum in solar wind plasma
in the context of compressible, driven-dissipative, anisotropic 3D MHD
turbulence.  Understanding turbulent cascades in the MHD plasma is
critical to many astrophysical phenomena. These range from
understanding the role of waves and nonlinear cascades in the
evolution of the solar wind, structure formation at the largest
scales, cosmic ray scattering and energization by solar wind
turbulence at the smallest scales and the heating of the solar wind
(Scalo \& Elmegreen 2004, Elmegreen \& Scalo 2004, Goldreich \&
Sridhar 1995, Zank 1999, Goldstein et al 1995, Scalo et al 1998,
Goldreich 2001) to problems such as energy transfer across many scales
in the ISM
\cite{roy,Haverkorn,ryu,rickett,Willett,elmegreen2,Dickey,minter}.

We find from our 3D (decaying turbulence) simulations that a
$k^{-5/3}$ density fluctuation spectrum emerges in fully developed
compressible MHD turbulence from nonlinear mode coupling interactions
that lead to the migration of spectral energy in the transverse
(i.e. ${\bf k} \perp {\bf U}$) Alfv\'enic fluctuations, while the
longitudinal ``compressional modes'' corresponding to ${\bf k}
\parallel {\bf U}$ fluctuations make an insignificant contribution to
the spectral transfer of inertial range turbulent energy. The
explanation, in part, resides with the evolutionary characteristics of
the MHD plasma that governs the evolution of the non-solenoidal
velocity field in the momentum field. It is the non-solenoidal
component of plasma motions that describes the high frequency
contribution corresponding to the acoustic time-scales in the modified
pseudosound relationship (Montgomery et al 1987; Matthaeus et al 1988;
Zank \& Matthaeus 1990; Zank \& Matthaeus 1993).  What is notable in
our present work is we find a self-consistent evolution of a
Kolmogorov-like density fluctuation spectrum in MHD turbulence that
results primarily from turbulent damping of non-solenoidal modes that
constitute fast and slow propagating magnetoacoustic compressional
perturbations. These are essentially a higher frequency (compared with
the Alfv\'enic waves) component that evolve on acoustic time-scales
and can lead to a ``pseudosound relationship'' as identified in the
nearly incompressible theory (Matthaeus et al 1988; Zank \& Matthaeus
1990; Zank \& Matthaeus 1993; Bayly et al; Shaikh \& Zank 2004a,b,c,
2006, 2007). The most significant point to emerge from our simulation
is the diminishing of the high frequency component that is related to
the damping of compressible plasma motion. This further leads to the
dissipation of the small scale and high frequency compressive
turbulent modes. Consequently, the MHD plasma relaxes towards a nearly
incompressible state where the density is convected passively by the
velocity field and eventually develops a $k^{-5/3}$ spectrum.  This
physical picture suggests that a nearly incompressible state develops
naturally from a compressive MHD magnetoplasma in the solar wind.

The higher resolution in-situ solar wind velocity field observations
at 1AU show a typical spectral index in the range from 1.4 to 1.5
(Podesta et al 2007; Tessein et al. 2009).  The latter clearly differs
from our simulations in which the spectrum of density field is close
to 1.67. To this end, we should point out that the solar wind
observations at 1 AU {\em contradict} the fundamental theme of our
model which states that the density fluctuations are convected
passively by the velocity field and thereby acquire similar spectral
power-law.  Owing thus to the differences between the observations and
our simulations, our model describing the passive convection of the
density field through the nearly incompressible velocity field may not
hold near 1 AU. Consequently, the solar wind density fluctuations
($\sim k^{-5/3}$) at 1 AU cannot be correlated with the velocity field
($\sim k^{-3/2}$) through the passive-advection phenomenology.  We
believe that the deviation of the density and velocity fields, leading
thereby to the obvious differences near 1 AU, may arise from a number
of physical processes, as noted below.  Firstly, the physical
processes driving the velocity spectrum near 1 AU are different from
those beyond 1 AU. Therefore, the evolutionary characteristics
associated with the velocity spectrum near 1 AU (describing a 3/2-like
spectrum) can certainly not be {\em representive} of the distant outer
heliospheric turbulence spectrum.  Secondly, close to 1 AU, stream
interactions, shear instabilities, compressional modes, and many other
physical processes can plausibly alter the velocity field spectrum. It
is unclear from the work of Podesta et al (2007) and Tessien et al
(2009) whether any of these processes have a significant influence on
the observed spectral indices.  To further clarify this point, we have
carried out more simulations to distinguish the evolution of kinetic
energy, and incompressible, compressible, and total velocity spectra
(not shown in this paper).  We find that the compressive velocity
field exhibits a flatter spectrum, while the incompressible velocity
field follows a 5/3 spectrum.  This could very well mean that the
velocity field, if is, driven or dominated by compressive modes, has a
spectrum may be a flatter than 5/3. Thirdly, the Alfv\'enic
interactions, often invoked to explain the descrepnacy between the
Kolmogorov-like (5/3) and Kraichnan-like (3/2) spectra, are ascribed
to Alfv\'en waves in MHD turbulence (Iroshnikov 1963; Kraichnan 1965;
Shebalin et al 1983; Biskamp 2003). The latter inhibits energy
cascades along the direction of propagation. The spectral transfer of
inertial range turbulent energy is therefore suppressed along the mean
$B_0$, whereas the perpendicular cascade is governed predominantly by
the hydrodynamic-like interactions. Consequently, the energy spectrum
is dominated by hydrodynamic-like processes which lead to a
Kolmogorov-like (5/3) inertial range turbulent spectrum. Hence a
Kolmogorov-like (5/3) spectrum emerges in MHD turbulence when the
nonlinear interactions are dominated by hydrodynamic-like eddies. By
contrast, magnetic field eddies governing the Alfv\'enic interactions
lead to a Kraichnan-like (3/2) spectrum in MHD turbulence.  Thus, the
disparate time scales associated with Alfv\'enic and compressive modes
may flatten out the velocity field but not the density field.  Our
model, thus relating the density to the velocity field spectrum, may
not hold near 1 AU in the circumstances where the velocity
fluctuations are driven by the processes as described above.

The Kolmogorov-like $k^{-5/3}$ spectrum in density, velocity and
magnetic fields resulting from our simulation (see Fig 1) are fully
consistent with those of Tilley \& Pudritz (2007), Mac Low et al
(1998, 1989), Biskamp (2003), Padoan \& Nordlund (1999), Stone et al
(1998), Goldstein et al (1995), Goldreich \& Sridhar (1995) and with
others that are described elsewhere in our paper.



 The support of NASA(NNG-05GH38) and NSF (ATM-0317509) grants is  acknowledged.




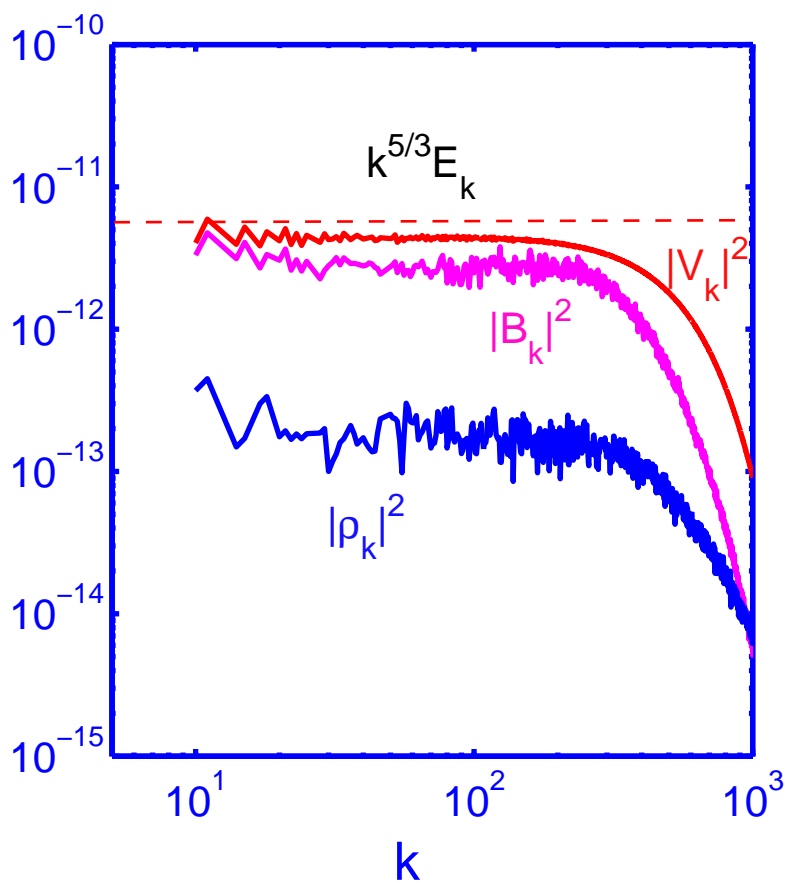
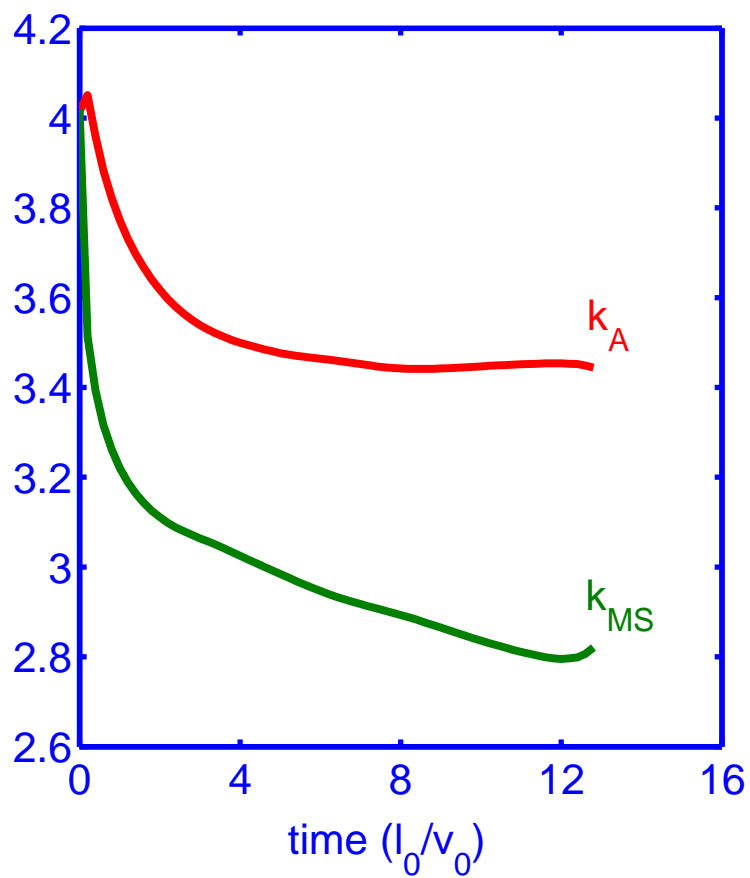
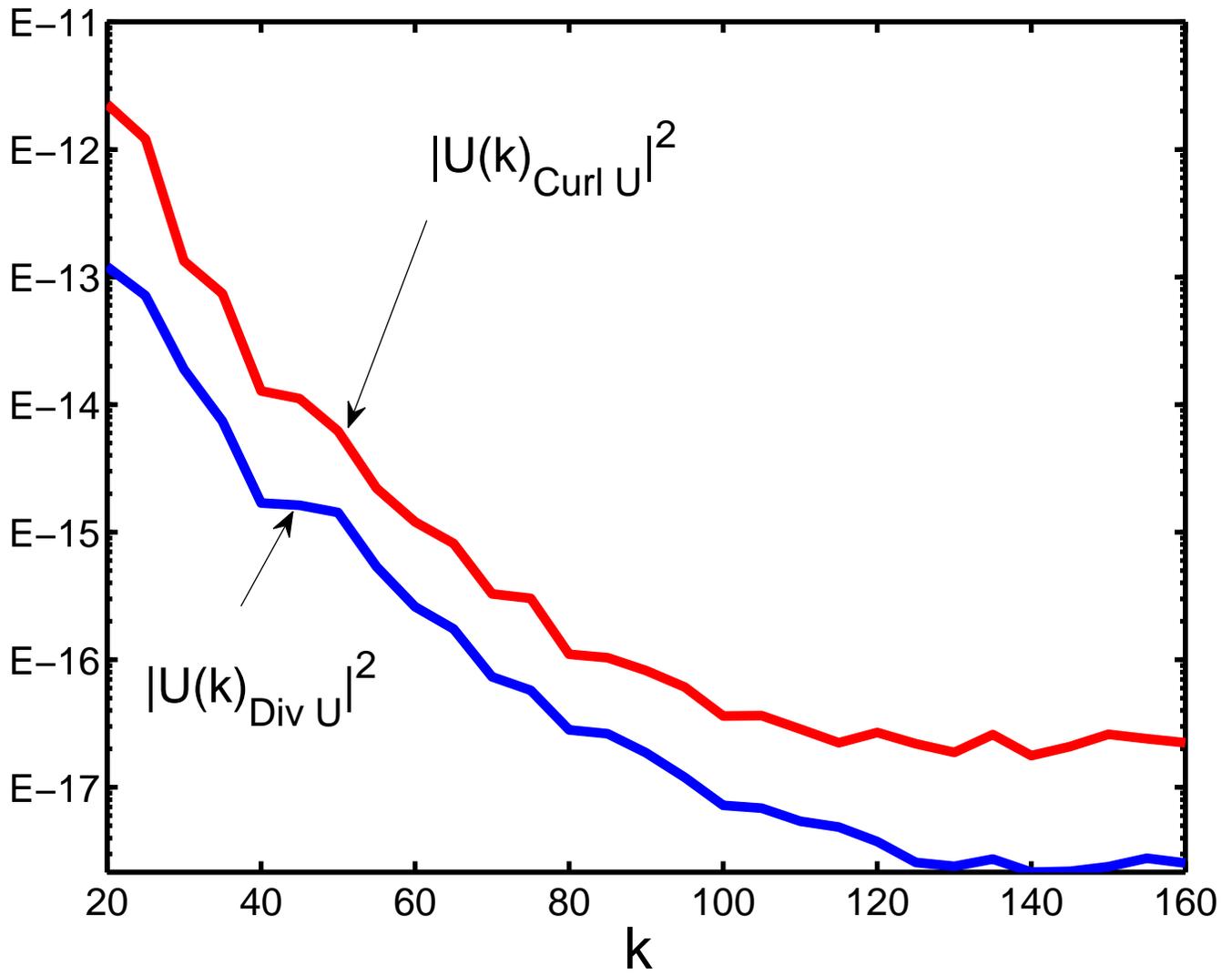
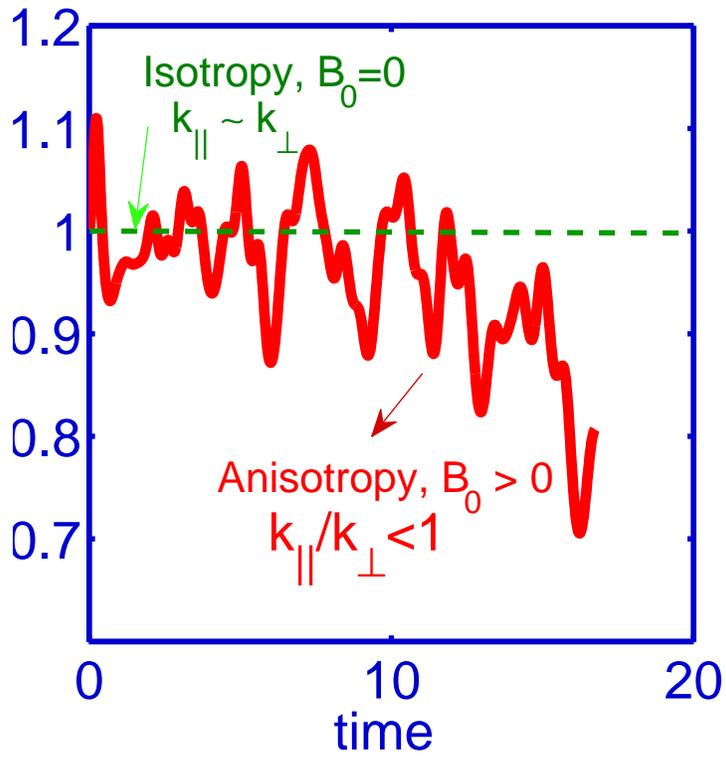
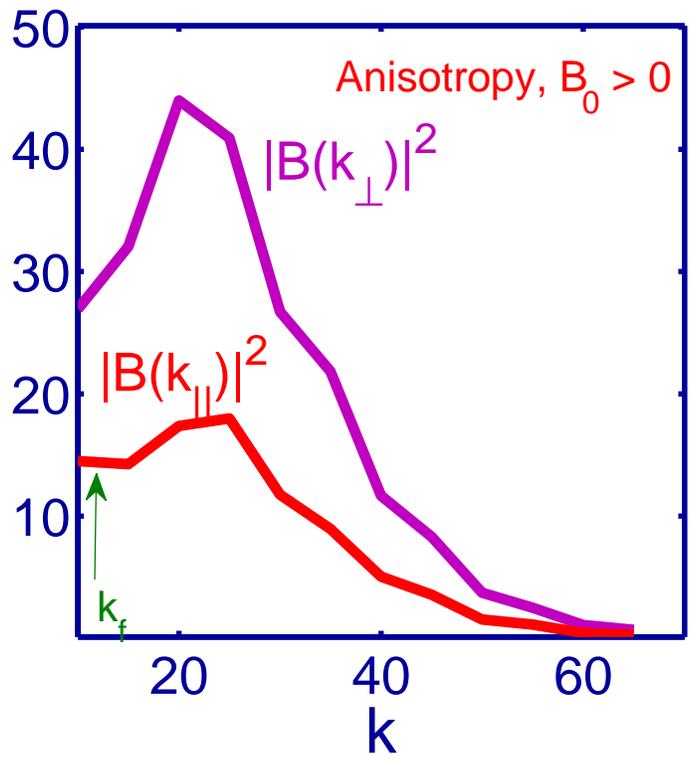
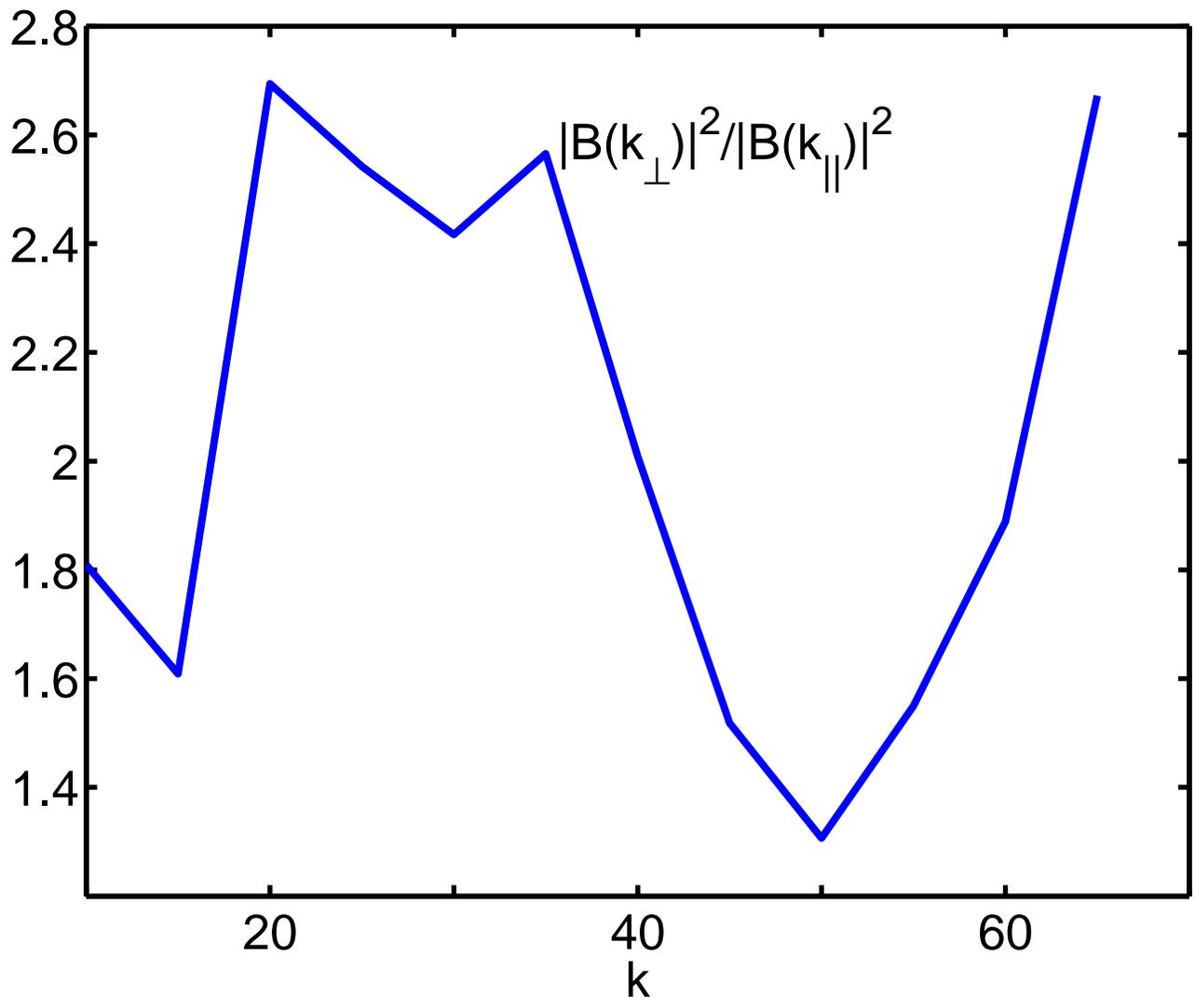
\end{document}